\documentclass[preprint,prx,longbibliography]{revtex4-1}
\usepackage{subcaption}
\usepackage{graphicx}
\usepackage{comment}
\usepackage{amsmath,amssymb,amsthm} 
\usepackage{scalerel,stackengine}
\usepackage{bm}
\usepackage{verbatim}
\usepackage{float}

\begin{document}
\title{Discrete Newtonian dynamics with  Nos\'{e}-Hoover thermostats}
\author{ S\o ren  Toxvaerd }
\affiliation{ Department
 of Science and Environment, Roskilde University, Postbox 260, DK-4000 Roskilde}
\email{st@ruc.dk}

\begin{abstract}
	Almost all  Molecular Dynamics (MD) simulations are  discrete dynamics with Newton's  algorithm first published in 1687, and much later by L. Verlet in 1967.
Discrete Newtonian dynamics has
the same qualities as Newton's classical analytic dynamics.  Verlet
	also published a first-order expression  for the instant 
temperature  which is inaccurate but presumably used in most MD simulations. One of the motivations for the present article is to
	correct this unnecessary inaccuracy in  $NVT$ MD dynamics. Another motivation is to derive simple algorithms for
	the Nos\'{e}-Hoover $NVT$ dynamics (NH) with the correct temperature constraint.
The simulations with NH discrete Newtonian dynamics show that the  NH works excellent for a wide range of the response time $\tau$
	of the NH thermostat, but
 NH simulations  favor a choice of a short response time, even shorter than the discrete time increment $\delta t$ used in  MD,
 to avoid large oscillations of the temperature.
\end{abstract}
\maketitle
\section{Introduction}
Newton published  PHILOSOPHI\AE \ NATURALIS PRINCIPIA MATHEMATICA ($Principia$) in 1687  \cite{Newton1687,Toxvaerd2023} 
and he began $Principia$
by  postulating the discrete Newtonian dynamics in \textit{Proposition I}, where he used the algorithm for
discrete dynamics
  to derive his second law for classical mechanics. The classical analytic dynamics of $N$ interacting objects
  is the exact solution, obtained by solving
  the coupled second-order analytic differential equations for the accelerations of the objects. The exact analytic solution 
  for the dynamical evolution of the positions of the objects is characterized by being time-reversible and
  symplectic, and by having three invariances for a conservative system: the conservation of
  momentum, angular momentum, and the energy of the system.
But Newton's discrete dynamics is also time reversible and symplectic and has  the same three invariances \cite{Toxvaerd2023} , and
  if the exactness of classical mechanics is given by these five qualities, then his discrete dynamics is also exact.
  Furthermore, there exists a shadow Hamiltonian
 where discrete  positions are located on the analytical trajectories for
 the shadow Hamiltonian \cite{Toxvaerd2023,Toxvaerd1994,Toxvaerd2012}. This means that there is no qualitative difference between
Newton's analytic and  discrete dynamics.

  The discrete Newtonian algorithm
	was much later in 1967
  rediscovered by Loop Verlet \cite{Verlet1967} and used
  in his pioneering Molecular Dynamics (MD) simulation of a system of Lennard-Jones particles. The algorithm has been
  reformulated several times and appears under a variety of names, the most well-known is the Leapfrog version.
	  Almost all MD simulations today in Natural Science  are performed by Newton's discrete algorithm. A review of the 
   discrete Newtonian dynamics is given in reference No.2 \cite{Toxvaerd2023}  with the proofs  of the invariances for a conservative system. The
  proof of the conservation of the energy is also given here in the appendix.

	Loop Verlet also derived  a first-order expression $\textbf{v}_{0,i}(t)$ for the discrete velocity  $\textbf{v}_{i}(t)$ 
	for object No. $i$ at time $t$  
obtained by a symmetric Taylor expansion \cite{Toxvaerd2023appendix}
\begin{equation}
	\textbf{v}_{i}(t)=\textbf{v}_{0,i}(t) -\frac{1}{6}\frac{\delta t^2}{m_i} \textbf{f}_{i}'(t)+ \mathcal{O}(\delta t^4\textbf{f}_{i}'''(t)) 
\end{equation}
with
	\begin{equation}
		\textbf{v}_{0,i}(t)=\frac{\textbf{r}_i(t+\delta t)- \textbf{r}_i(t-\delta t)}{2 \delta t}
	\end{equation}
  from its positions $ \textbf{r}_i(t-\delta t)$ and $\textbf{r}_i(t+\delta t)$, and  this expression is used in today's MD \cite{Tildesley,Frenkel}.
 The velocity $\textbf{v}_{0,i}(t)$ in Eq. (2) is   the first term for the velocity  in Verlet's forward and backward Taylor
 expansion of the discrete position $\textbf{r}(t)$, 
and if the velocity is approximated by this expression, it entails a systematic and sometimes significant error in the calculated
temperature and in the heat capacity of a system \cite{Toxvaerd2023a}. It is the one of the reasons
		for this article, where the correct expression for the velocity and the kinetic energy is used to
		revise the formulation of the $NVT$ Nos\'{e}-Hoover algorithm (NH) for  $NVT$  simulations \cite{Hoover1985}. 
Constant temperature MD has been simulated for many decades and the method is reviewed in \cite{Harish2021}. Many of the algorithms for discrete dynamics at constant temperatures
are, however, with broken time symmetry and  irreversible dynamics. However, the NH dynamics, which is derived from the steady state Lioville equation for
	an extended phase space with  friction with a friction coefficient $\eta(t)$, is time reversible \cite{Hoover1985,Hoover1991}.

	The article starts in Section 2 by presenting the expressions for the velocities, kinetic energy, and  temperature by the use
	of the energy invariance derived in the appendix, and  the NH  algorithms for $NVT$ discrete dynamics are presented
in Section 3.
	The algorithms are used in Section 4 to obtain $NVT$ state points for a system of Lennard-Jones particles (LJ). Section 5 summarizes
	the results.

\section{Discrete Newtonian dynamics}
  	
  The classical discrete dynamics between $N$ spherically symmetrical objects
     with masses $ m^N=m_1, m_2,..m_i,..,m_N$ and positions \textbf{r}$^N(t)=$\textbf{r}$_1(t)$, \textbf{r}$_2(t)
     ,..,$\textbf{r}$_i(t),..$\textbf{r}$_N(t)$  is obtained 
 by Newton's discrete algorithm. Let the force, $ \textbf{f}_i$ on object No. $i$ be a sum of pairwise  forces  $ \textbf{f}_{ij}$ between pairs of   objects $i$ and $j$
 \begin{equation}
	 \textbf{f}_i=  \sum_{j \neq i}^{N} \textbf{f}_{ij}.
 \end{equation}	
  Newton's discrete algorithm 
	is a  symmetrical time-centered difference whereby the dynamics is time reversible and symplectic.
The Verlet formulation of the algorithm is
\begin{equation}
	\textbf{r}_i(t+ \delta t)=2 \textbf{r}_i(t)-	\textbf{r}_i(t- \delta t)+ \frac{\delta t^2}{m_i}\textbf{f}_i(i).
\end{equation}

 The  energy in analytic dynamics is the sum of potential energy $ U(\textbf{r}^N(t))$ and kinetic energy $K(t)$, 
 and it is an invariance for a conservative  system with analytic dynamics.
 There is, however, a problem with the determination of the kinetic energy in the discrete dynamics since the velocities at time $t$ are not known. 
Traditionally one uses Eq. (2) and the expressions
\begin{eqnarray}
	K_0(t)=  \sum_i^N \frac{1}{2}m_i \textbf{v}_{0,i}(t)^2 \\
	E_0(t)= U(\textbf{r}^N(t))	+K_0(t)
 \end{eqnarray}	
 for the velocity, kinetic energy $K(t)$, potential energy $ U(\textbf{r}^N(t))$ and energy $E(t)$ in MD. But 
  the total energy $E_0$ obtained by using Eq. (6) with $K(t)=K_0(t)$ for the kinetic energy 
  fluctuates with time although it remains constant, averaged over long time intervals.
 This is due to the fundamental quality of Newton's discrete dynamics, where the positions and
momenta appear asynchronous and with a discrete change in momentum at time $t$. The energy invariance in discrete Newtonian dynamics (D)
is derived in the appendix. The kinetic energy $K_D(t)$ in the time interval $[ t-\delta t/2,  t+\delta t/2]$ is
\begin{eqnarray}
	K_{\textrm{D}}=  \frac{1}{2}( K_{\textrm{D}_+}+ K_{\textrm{D}_-})=                         \nonumber \\
=\frac{1}{2}\sum_i^N\frac{1}{2}m_i [\frac{\textbf(\textbf{r}_i(t+\delta t)-\textbf{r}_i(t))^2}{\delta t^2}+
\frac{\textbf(\textbf{r}_i(t)-\textbf{r}_i(t-\delta t))^2}{\delta t^2}]\nonumber \\
	=\frac{1}{2}\sum_i^N\frac{1}{2}m_i(\textbf{v}_i(t+\delta t/2)^2+\textbf{v}_i(t-\delta t/2)^2).
 \end{eqnarray}

\subsection{The  temperature in discrete Newtonian dynamics }
 The constant kinetic energies in the time intervals $[t-\delta t/2,t]$ and  $[t,t+\delta t/2]$  in between a force impulse  at time $t$ are related  \cite{Toxvaerd2013}.
 It is easy to derive the relation
\begin{equation}
	\textbf{v}_{0,i}(t)^2=\frac{1}{2}\textbf{v}_i(t+\delta t/2)^2+\frac{1}{2}\textbf{v}_i(t-\delta t/2)^2- \frac{1}{4}(\delta t/ m_i \textbf{f}_i(t))^2
\end{equation}
 from Eq. (4) between the  first-order expression  Eq. (2) for the square of the  velocity at time $t$  and the
well-defined expression for the  square of the  velocities $\textbf{v}_i(t-\delta t/2)$ and  $\textbf{v}_i(t+\delta t/2)$  in the time intervals $[t-\delta t/2, t]$ and  $[t, t+\delta t/2]$.
The corresponding expression for the kinetic energy, $K_0(t)$, and the traditional value for the temperature used in MD simulations
\begin{equation}
	k_{\textrm{B}}T_0(t)=\frac{<2 K_0(t)>}{N_f}
\end{equation}	
 used in MD simulations
	for a system with $N_f$ degrees of freedom is less than the  mean kinetic energy, $K_D(t)$, and the temperature
\begin{equation}
	k_{\textrm{B}}T_{\textrm{D}}(t)= \frac{<2 K_D(t)>}{N_f}.
\end{equation}	
In the
discrete time interval   $[t-\delta t/2, t+\delta t/2]$  the relation is
\begin{equation}
 K_0(t)=K_D(t)-\sum_i^N \frac{1}{8}\delta t^2/ m_i \textbf{f}_i(t)^2.
\end{equation}

The correct temperature $T_D$ is bigger than $T_0$ and the difference between $T_D$ and $T_0$ 
increases with  density, temperature, and the discrete time step $\delta t$. The difference is
relatively small and of the order of a few percent at state points with low pressure, but  it is significant for state points
with high densities, pressure,  temperatures,  the strength of the repulsive forces, or for large time increments $\delta t$ \cite{Toxvaerd2023a}.

\section{ \textit{NVT} ensemble simulations with the Nos\'{e}-Hoover thermostats}

  Many \textit{NVT} ensemble simulations in MD are   with the Nos\'{e}-Hoover thermostat (NH-MD)
  where a friction  $\eta(t)\textbf{p}_i(t)$ 
  acts simultaneously with the force $\textbf{f}_i(t)$ \cite{Hoover1985}.
 The   Hamilton  formulation
of classical dynamics with the friction is \cite{Hoover1985,Hoover1991}
\begin{eqnarray}
 \dot {\textbf{r}}_i(t)= \textbf{p}_i(t)\\
\dot {\textbf{p}}_i(t)= \textbf{f}_i(t)/m_i-\eta(t)\textbf{p}_i(t)\\
\dot{\eta}(t)=\alpha^{-1}(K(t)-K),
\end{eqnarray}
with a  restoring  friction  $\eta(t)\textbf{p}_i(t)$  which constrains the kinetic energy $K(t)$  to  $K =2N_fk_{\textrm{B}}T$.
The damping factor $\alpha$ is expressed as
\begin{equation}
 \alpha= N_f \tau/2,
\end{equation} where $\tau$ is a characteristic ''response" time of the thermostat  \cite{Toxvaerd1991}.

The discrete Newtonian dynamics  with the time symmetrical Leapfrog formulation  with NH-MD is
\begin{eqnarray}
\textbf{r}_i(t+\delta t)= \textbf{r}_i(t)+ \delta t \textbf{v}_i(t+\delta t/2) \\
	\textbf{v}_i(t+\delta t/2)=  \textbf{v}_i(t-\delta t/2)+ \delta t/m_i  \textbf{f}_i(t) \nonumber \\
	-\eta(t)(\textbf{v}_i(t+\delta t/2)+ \textbf{v}_i(t-\delta t/2))/2, 
\end{eqnarray}
and the discrete friction variable $\eta(t)$ can either be 
 updated from two sets of discrete values $\eta(t-\delta t), \eta(t)$ by
\begin{equation}
	\eta(t+\delta t)=\eta(t- \delta t)+ \delta t/\tau(T_D(t)- T)  \textrm{  Method I},
\end{equation}
or by
\begin{equation}
	\eta(t+\delta t)= \eta(t)+ \delta t/\tau(T_{D_+}(t)- T)  \textrm{  Method  II},
\end{equation}
where $k_{\textrm{B}}T_{D_+}=  \sum_i^N  m_i \textbf{v}(t+\delta t/2)^2/N_f $.
 Eq. (18)
is the time centred analog to Newton's discrete central difference algorithm.
The time symmetry and time
reversibility is maintained in the NH-MD   discrete dynamics with both methods,
and both methods have been used in NH-MD \cite{Toxvaerd1991,Hoover1990}.

The NH dynamics can be extended to include more than one thermostat and this has been used in simulations of systems with organic  chain molecules.
 The thermostats can equilibrate specific modes in the systems or act on all $N_f$ degrees of freedom
\cite{Frenkelchap6}. 
In the present formulation of NH dynamics and with $n$ thermostats which act with all $N_f$ modes

\begin{eqnarray}
		\eta_1(t+\delta t)=\eta_1(t- \delta t)+ \delta t/\tau_1(T_D(t)- T) \nonumber \\
	\eta_2(t+\delta t)=\eta_2(t- \delta t)+ \delta t/\tau_2(T_D(t)- T) \nonumber \\
	.........,	
\end{eqnarray}
and with a total friction $\eta= \Sigma_i^n \eta_i/n$.

Eq. (17) can be rearranged to
\begin{equation}
	\textbf{v}_i(t+\delta t/2)=  \frac{ \textbf{v}_i(t-\delta t/2)(1-\eta(t)/2)+  \delta t/m_i  \textbf{f}_i(t)}
	{ 1+\eta(t)/2},
\end{equation}
and the  $NVT$  time reversible NH-MD  dynamics is obtained by the  NH-MD algorithm: Eqn. (16) and (21) with either Eq. (18), and/or Eq. (19)
, and/or Eq. (20).

	 \begin{figure}
	  \begin{center}	  
 	 \includegraphics[width=8.6cm,angle=-90]{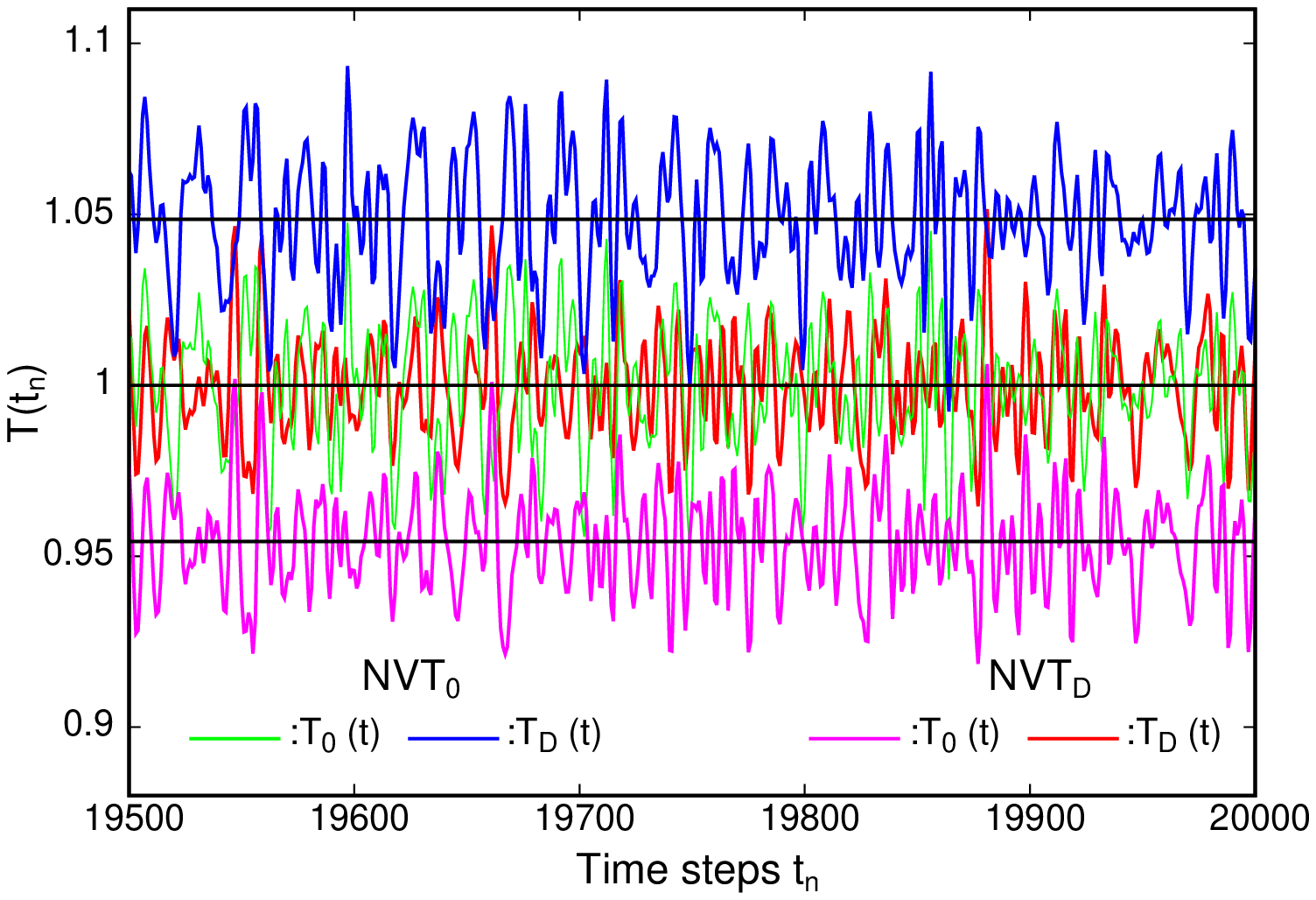}
		  \caption{NH-MD with $T_{Thermostat}=T_0=1.00$ $(NVT_0$, old NH-MD), and with  $T_{Thermostat}=T_D=1.00$ $(NVT_D$, present NH-MD). 
		  The figure shows five hundred  (representative) time steps for the simulations of $10^6$ time steps.
		  The temperatures $\bar{T}_D$ and the root mean square (rms) fluctuations are given in Table I. }
  \end{center}		  
  \end{figure}
	 \begin{figure}
	  \begin{center}	  
 	 \includegraphics[width=8.6cm,angle=-90]{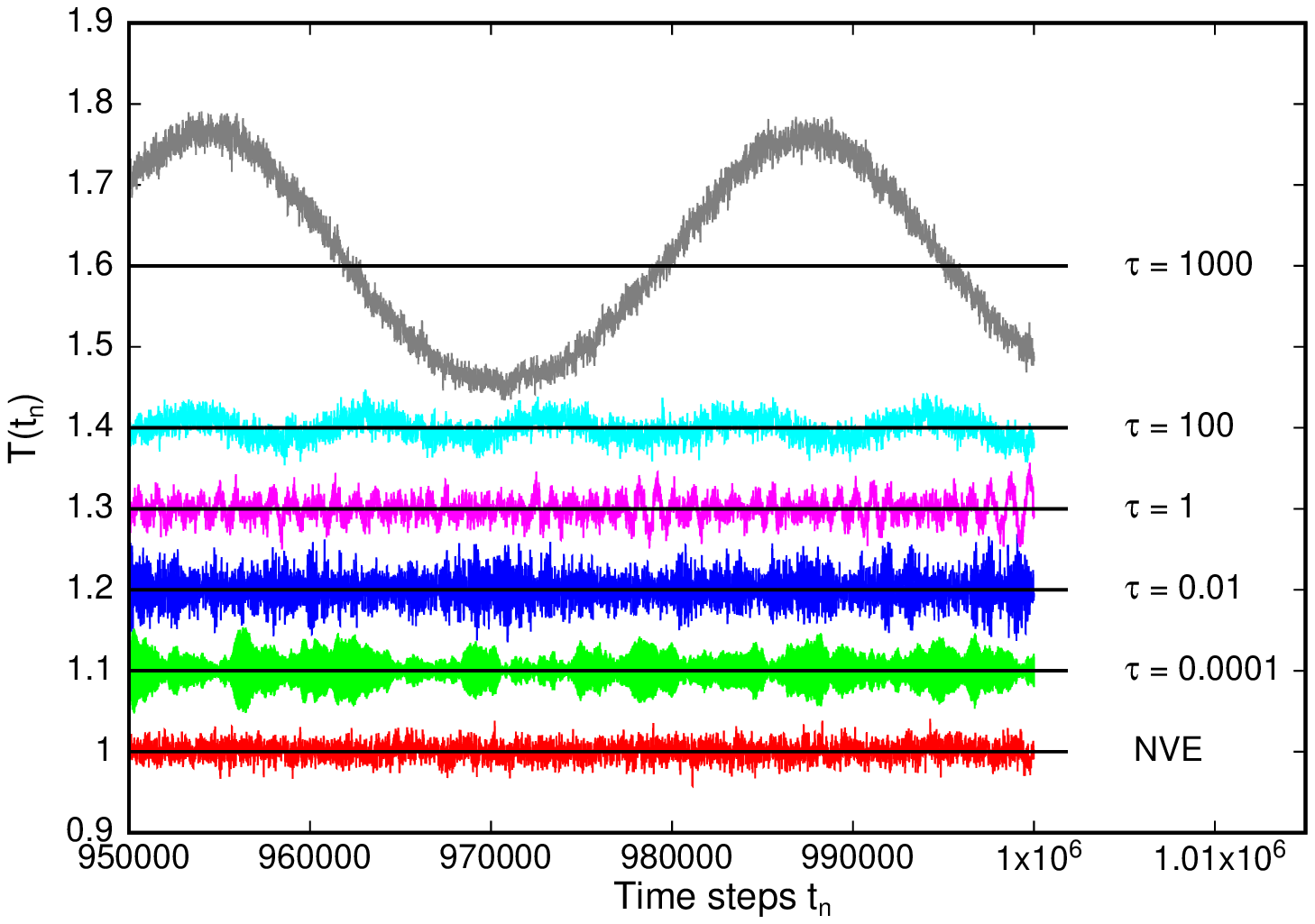}
		  \caption{Five evolutions of $T_D(t)$  in a LJ system of $N=2000$ particles  with 
		   $\delta t=0.005$ and with a NH thermostat with $T=1.0$ (Method I, Eq. (18)) together with the corresponding
		   temperature without a thermostat (NVE). (Four NH-evolutions are parallel shifted with succeding $\Delta T=0.1$, and the last
		   evolution with $\tau=1000$ is shifted with  $\Delta T=0.2$.) All six simulations are started
		  from the same  NVE start configuration with different values of $\tau$. The figure shows the last 50000  time steps
		  for the simulations of $10^6$ time steps. The temperatures $\bar{T}_D$ and the root mean square (rms) fluctuations are given in Table I. }
  \end{center}		  
  \end{figure}

\begin{table}
	\caption{\label{Table I} Temperatures in the LJ system.}
	\centering
	\begin{tabular}{c c c c c}
		$ \tau $ & $T_D \pm rms $  & $ T_0 \pm rms$  & $ \eta \pm  rms $ & Method \\ 

 \hline
			  -    &  1.000000000 $ \pm 9.5\times10^{-3} $ & -  & NVE &   \\
			  0.0001    &    0.999999993 $ \pm 1.7\times 10^{-2 }$ & 0.999014   $ \pm 1.7\times 10^{-2 }$ &  $ -2.06\times10^{-6} \pm  0.18 $   &  I   \\
			  0.0001    &   1.000000025  $ \pm 1.8\times 10^{-2 }$ &  0.999023    $ \pm 1.8\times 10^{-2 }$ &   $ 1.7\times10^{-6} \pm  1.4 \times10^{-2} $   &  II   \\
			  0.001 &  1.000000970$\pm 1.7 \times 10^{-2 }$  &   0.999025  $ \pm 1.7\times 10^{-2 }$ & $ -2.81\times 10^{-6 } \pm 2.1 \times 10^{-3}$& I\\
			  0.01 & 1.000000052$\pm 1.8 \times 10^{-2 }$  &  0.999024  $ \pm 1.8\times 10^{-2 }$ & $ $ $ 2.39\times 10^{-6 } \pm 2.1 \times 0.1277$& I\\
			  0.1 &   0.999998637$\pm 1.8 \times 10^{-2 }$  &  0.999023  $ \pm 1.8\times 10^{-2 }$ & $ -1.68\times 10^{-6 } \pm 4.0 \times10^{-2}$ & I\\
			  1&  1.000005194  $\pm 1.8 \times 10^{-2 }$  &  0.999030  $ \pm 1.8\times 10^{-2 }$ & $ $ $ 3.76 \times 10^{-7 } \pm 1.3 \times10^{-2}$ & I\\
			  1&   0.999995501   $\pm 1.9 \times 10^{-2 }$   &   0.999029  $ \pm 1.9\times 10^{-2 }$ & $ -9.16 \times 10^{-7 } \pm 3.2 \times10^{-3}$ & II\\
			  10&   0.999986336  $\pm 1.9 \times 10^{-2 }$  &  0.999011  $ \pm 1.7\times 10^{-2 }$ & $ $ $ 1.29 \times 10^{-6 } \pm 4.3 \times10^{-3}$  & I\\
			  100&  0.999883749    $\pm 2.9 \times 10^{-2 }$  & 0.998908  $ \pm 2.9\times 10^{-2 }$ & $ $ $ 2.3 \times 10^{-6 } \pm 2.3 \times10^{-3} $ & I\\
			  1000&   0.999458904    $\pm 0.12108 $  &  0.998477   $ \pm  0.12083 $ &  $ 1.73 \times 10^{-5 } \pm 3.2 \times10^{-3}$  & I\\
			  1000&    0.999516366     $\pm 0.1236 $  &   & $ $ $ 1.50 \times 10^{-5 } \pm 3.2 \times10^{-3}$  & II
\end{tabular}
\end{table}

	 \begin{figure}
	  \begin{center}	  
 	 \includegraphics[width=8.6cm,angle=-90]{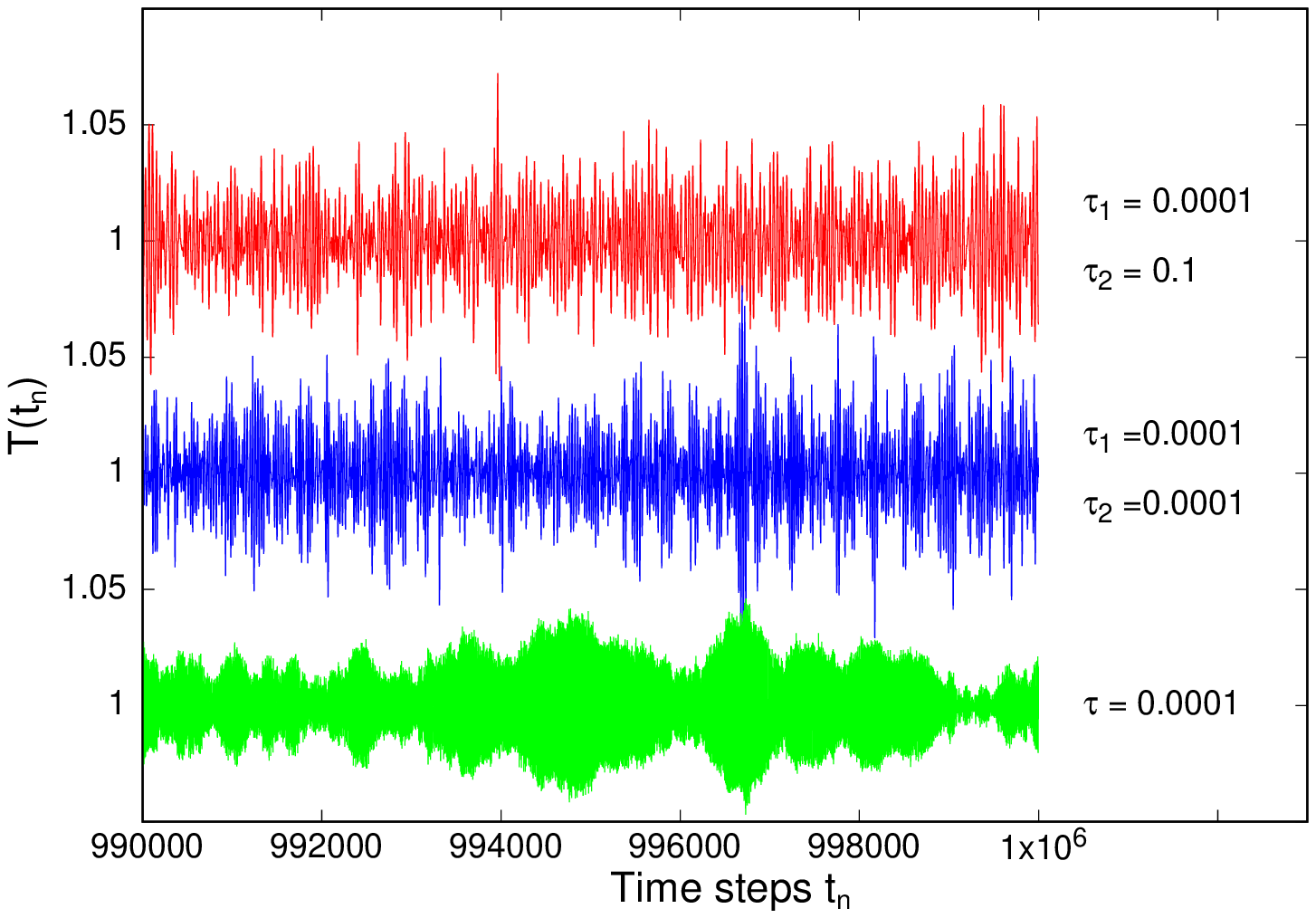}
		  \caption{ The temperature fluctuations of $T_D(t)$ in the
		  temperature interval [0.95, 1.05] for NH-dynamics with  one  and with two thermostats.
		  The fluctuations  of the temperature with one thermostat and 
		  $\tau=0.0001$ are also shown in green in Figure 1. The fluctuations in the temperature interval [0.95, 1.05]
		  with two thermostates are shown above, and for two sets of frictions.
		  With red is for $\tau_1=0.0001$ and $\tau_2=0.1$. 
		  With blue is  $T_D(t)$  with  $\tau_1=\tau_2=0.0001$ with different start values of $\eta_1(0), \eta_{1}(-\delta t)$ and 
		  $\eta_2(0), \eta_2(-\delta t)$, respectively. 
		  }
	  \end{center}		  
  \end{figure}

	 \begin{figure}
	  \begin{center}	  
 	 \includegraphics[width=8.6cm,angle=-90]{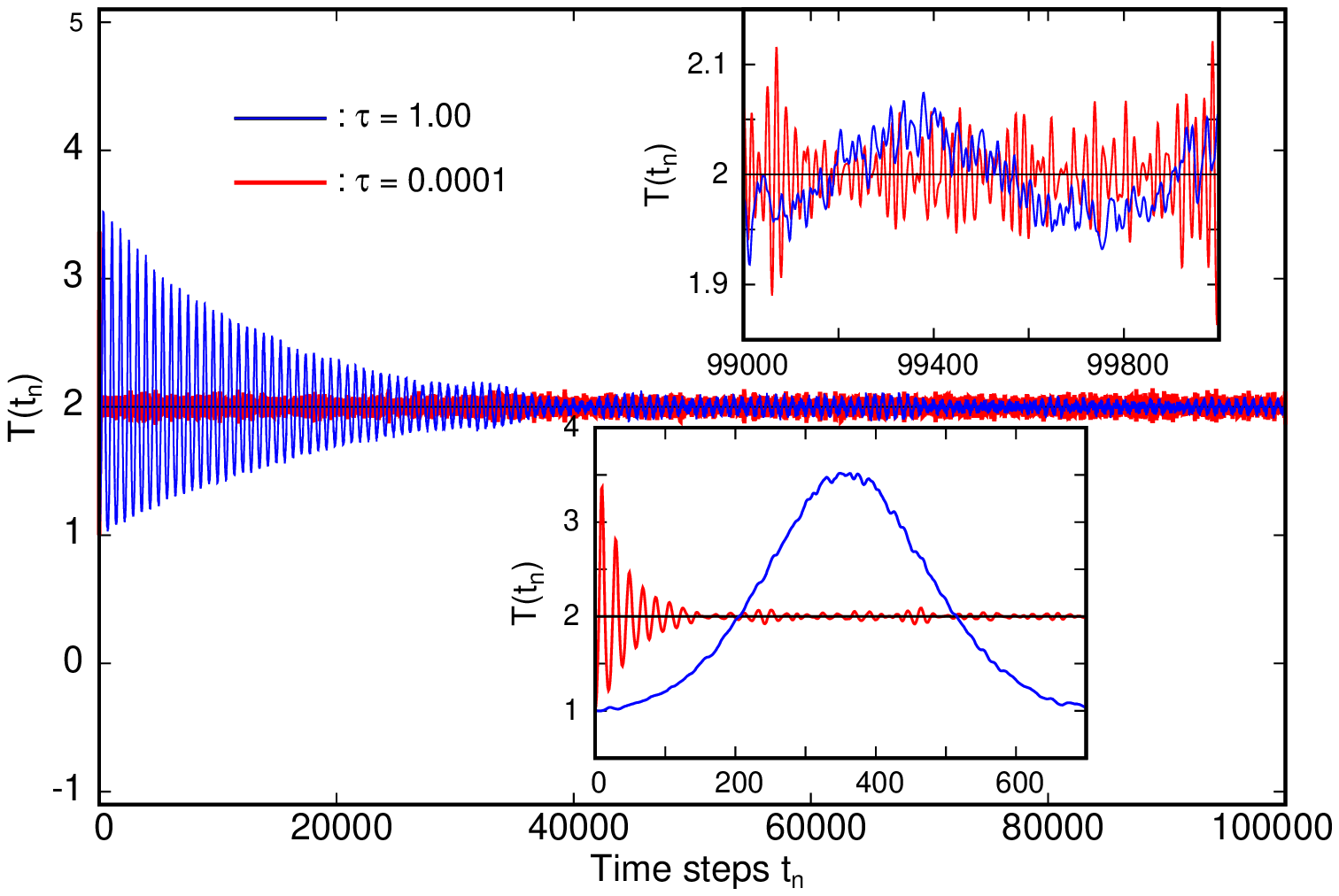}
		  \caption{Temperature evolution in the LJ system   by changing the thermostat temperature from $T_1=1.00$ to $T_2=2.00$.
		  With blue is $T_D(t) $ for a relative long response time $\tau=1.00$ and with red is for a short response time  $\tau=0.0001$.
		  The lower inset shows the temperature evolution at the start of the calibration and
		  the upper inset is at the end of the calibration of the temperature. 
		  }
	  \end{center}		  
  \end{figure}

  \section{$NVT$ simulations}

  The Nos\'{e}-Hoover friction $\eta(t)$ thermostats the system's instantaneous kinetic energy and temperature $T(t)$ using
kinetic energy, Eq. (18) or Eq.(19). Most $NVT_0$ simulations with NH-MD use Verlet's expression $T_0(t)$, Eq. (9), for the temperature.
The exact expression for the kinetic energy and the instant temperature  $T_D(t)$ by the discrete dynamics is, however systematically higher and
given by Eq. (10). The difference between the two canonical simulations, $NVT_0$, and $NVT_D$, increases with the discrete time increment
$\delta t$ and the force according to Eq. (8). It is of the order a few percent or less for many MD $NVE$ simulations at moderate
densities  and force fields \cite{Toxvaerd2023a}, but can be much larger for stronger forces.

 The differences between the two  ensemble simulations,  $NVT_0$ and $NVT_D$, where the instant temperature is either given by  $T_0(t)$
 or by  $T_D(t)$,  are determined for a  Lennard-Jones system (LJ). The Lennard-Jones system is a simple model of a system with the
 classical gas-liquid-solid thermodynamic behavior. Systems of  $N=2000$ LJ particles
 were simulated for various state points $\rho,T$. 
 The differences between  $NVT_0$ and $NVT_D$ 
 at the high density $[\rho=1.40, T=1.00]$ fluid state point, and with  $\delta t=0.010$ and $\tau=0.0010$  \cite{units} are shown in Figure 1.
 The NH thermostat constrains a specified kinetic energy either
 to a given temperature $T$ by Eq. (9): $NVT_0$ with $T_0(t)=T$, or  by Eq. (10):  $NVT_D$ with $T_D(t)=T$.  The mean canonical temperatures
 $\bar{T}_0$ and $\bar{T}_D$, respectively, are in both cases equal to the input temperature $T=1$, and in both cases the difference between
 $NVT_0$ and  $NVT_D$, obtained from $10^6$ time steps  are $\approx$ 4- 5 per cent.
 For $NVT_0$: $T_0=1.0000000$, green curve in Figure 1;  $T_D=1.049$, blue curve. 
 For $NVT_D$: $T_D=1.0000000$, red curve in Figure 1;  $T_0=0.954$, magenta curve. (The figure shows the difference in a short time interval of five hundred steps).

\subsection{ Nos\'{e}-Hoover  simulations of a Lennard-Jones system}
 The two
NH algorithms for $NVT$ dynamics ( I and II) are tested by MD simulations of a LJ system with the thermostats at various state points in the fluid state
 of the LJ system. The
first examples are for a liquid state point $(T, \rho)=(1, 0.60)$ near the coexisting liquid with the density $\rho_l=0.589$ at $T=1.00$ \cite{Watanabe2012}.
The temperatures $T_D(t)$ for various values of the NH-response time $\tau$ are shown in Figure 2, and the mean temperatures and their root mean
fluctuations are given in Table I. The thermostats work excellent for a wide range of the values of the response time $\tau$ but the system's
instant temperature $T_D(t)$ exhibits oscillations for values of the response time $\tau >> 1$, whereas the thermostat works excellent for smaller
values, and even for response times $\tau < \delta t$. 

Table I contains also   data for NH dynamics with Method II for $\tau=0.0001, 1$ and 1000, respectively,
and there are no differences between the  mean temperatures obtained by the two algorithms. The thermostats were tested for different
temperatures and densities, including a fluid system at the high density $\rho=1.40$
and the NH dynamics work excellently for both methods. The table for the $NVT_D$ with $T_D=1.00$  also lists the  mean values $\bar{T}_0$. They
 deviate  marginally from  $\bar{T}_D$ at this state point with pressure $P \approx 0$ near a liquid in coexisting with gas where the LJ particles
are near the minimum potential energy state, and where the forces are $\approx 0$.

The instant temperature $T_D(t)$ with $NVT_D$ dynamics, and for the
  smallest value of $\tau=0.0001$ is shown in green in Figure 2. The temperatures exhibit some
  ''zones" with different amplitudes of the fluctuations. The observed ''zones'' with a single thermostat friction
  can be caused by a nonergodicity of the NH-dynamics, and it can be removed by an additional  friction  \cite{Martyna1992}.
  The simulations of the LJ system at $T,\rho=1,0.6$ with two independent NH thermostats are shown in Figure 3. The fluctuation zones are removed
  by the use of two independent thermostats. However, extended investigations of NH-MD of systems with this nonergoditisity show, that there are systems
  for which it  is not possible to achieve ergodisity within computationally achievable time by the use of two NH thermostates \cite{Patra2014}.

  A typical situation in NH-MD is a calibration
  of a system from  a  temperature  $T_D(0)=T_1$ to another temperature $T_2$.
  The temperature evolution in the NH thermostated LJ system by changing the thermostat temperature from $T_1=1.00$ to $T_2=2.00$ is shown in
  Figure 4 for two different values of the response time $\tau$. With blue is $T_D(t)$ for  for a relative long response time $\tau=1.00$ and
 with red is  for a short response time  $\tau=0.0001$. The instant temperature oscillates in both cases around the new target temperature $T_2=2.00$,
 but the system with the short response time converges rapidly to instant temperatures near $T_2$ whereas the temperature oscillates with large amplitudes
 for many thousand time steps in the case with a long response time. The lower inset shows the evolution shortly after the start of the calibration
 and the upper inset are for the temperatures after $\approx$ hundred thousand time steps.

 \section{Summary}
 Almost all MD simulations are with Newtonian discrete dynamics with Newton's Verlet- or the equivalent Leapfrog algorithm,
and the discrete Newtonian dynamics has
the same qualities as Newton's classical analytic dynamics \cite{Toxvaerd2023}. Verlet also proposed an approximative  expression, Eq. (2),
for the  
temperature at $t_n$ after n time steps  \cite{Verlet1967}, which  presumably is used in most MD simulations, but the first-order expression is not useful at positions
where an object is exposed to a strong force e.g. at an ellastic collision.
The expression for the velocity at the discrete dynamics has been improved by including higher order terms in the Taylor expansion, Eq. (1),
by a spline fit to the discrete positions, or by using a higher-order predictor-corrector algorithm.
A. Rahman used a higher-order predictor-corrector 
algorithm in the very first MD simulation \cite{Rahman1964},
and also the Runge-Kutta algorithm is occasionally used in $NVT$ MD dynamics \cite{Orel1993}.
But almost all $NVT$ dynamics are with Newton's discrete algorithm,
a thermostat, and by using Eq. (2) for the velocity at $t_n$.

 The approximative  expressions for the velocity and kinetic energy at the time step $t_n$ can be
 replaced by the  exact expressions for velocities and kinetic energies in the time intervals at   $t_n$.
 One of the motivations for the present article is to
correct this unnecessary error in the determination  of the temperature in  Newton's discrete dynamics.
Another motivation is to derive simple algorithms for the Nos\'{e}-Hoover $NVT$ dynamics with the
correct kinetic constraint on the discrete Newtonian dynamics.

The simulations with NH-MD discrete Newtonian dynamics show that  NH-MD  works excellent for a wide range of the response time $\tau$, but
the NH-MD  favor a choice of a short response time, even shorter than the discrete time increment $\delta t$ in the MD,
in order to avoid large oscillations of the temperature.

\section{Appendix}

\textbf{The energy invariance}

The energy invariance, $E_{\textrm{D}}$
in Newton's discrete dynamics (D) can be seen by considering the change in kinetic energy, $\delta K_ {\textrm{D}},$
potential energy,  $\delta U_ {\textrm{D}},$ and
the work,  $W_{\textrm{D}}$ done by the force 
in the time interval $[t-\delta t/2, t+\delta t/2].$ The loss in  potential energy, $-\delta U_{\textrm{D}}$ is defined as
the work done by the forces at a  move of the positions \cite{Goldstein}. 
An expression for the work, $W_{\textrm{D}}$ done in the time interval by the discrete dynamics
from the position  $(\textbf{r}_i(t)+ (\textbf{r}_i(t-\delta t))/2$ at $t-\delta t/2$
to the position  $(\textbf{r}_i(t+\delta t)+ \textbf{r}_i(t))/2$ at $t+\delta t/2$ with the change in position
$\textbf(\textbf{r}_i(t+\delta t) -\textbf{r}_i(t-\delta t))/2$ is  \cite{Toxvaerd2023}
\begin{equation}
	-\delta U_{\textrm{D}}=W_{\textrm{D}}= \sum_i^N  \textbf{f}_i(t)(\textbf{r}_i(t+\delta t) -\textbf{r}_i(t-\delta t))/2.
\end{equation}	
By rewriting Eq. (4) to
\begin{equation}
	\textbf{r}_i(t+ \delta t) -\textbf{r}_i(t-\delta t)= 2(\textbf{r}_i(t) -\textbf{r}_i(t-\delta t))+\frac{\delta t^2}{m_i} \textbf{f}_i(t),
\end{equation}
and inserting in Eq. (22) one obtains an expression for the total work in the time interval
\begin{equation}
	-\delta U_{\textrm{D}}=  W_{\textrm{D}}	=  \sum_i^N  [(\textbf{r}_i(t) -\textbf{r}_i(t-\delta t)) \textbf{f}_i(t)+ \frac{\delta t^2}{2m_i}\textbf{f}_i^2].
\end{equation}

An expression for the mean kinetic energy $K_{\textrm{D}}$ of the discrete dynamics  in the time interval $[t-\delta t/2, t+\delta t/2]$ is

\begin{eqnarray}	
 K_{\textrm{D}}=   K_{\textrm{D}_+}+ K_{\textrm{D}_-}=                         \nonumber \\		
\frac{1}{2} \sum_i^N \frac{1}{2}m_i[\frac{\textbf(\textbf{r}_i(t+\delta t/2)-\textbf{r}_i(t))^2}{\delta (t/2)^2}+
\frac{\textbf(\textbf{r}_i(t)-\textbf{r}_i(t-\delta t/2))^2}{\delta (t/2)^2}] \nonumber \\
=\frac{1}{2}\sum_i^N\frac{1}{2}m_i [\frac{\textbf(\textbf{r}_i(t+\delta t)-\textbf{r}_i(t))^2}{\delta t^2}+
\frac{\textbf(\textbf{r}_i(t)-\textbf{r}_i(t-\delta t))^2}{\delta t^2}].
\end{eqnarray},
and with the change in the time interval
\begin{eqnarray}	
\delta K_{\textrm{D}}=   K_{\textrm{D}_+}- K_{\textrm{D}_-}                           \nonumber \\		
=\sum_i^N\frac{1}{2}m_i [\frac{\textbf(\textbf{r}_i(t+\delta t)-\textbf{r}_i(t))^2}{\delta t^2}-
\frac{\textbf(\textbf{r}_i(t)-\textbf{r}_i(t-\delta t))^2}{\delta t^2}].
\end{eqnarray},

By rewriting  Eq. (4) to
\begin{equation}
	\textbf{r}_i(t+ \delta t) -\textbf{r}_i(t)= \textbf{r}_i(t) -\textbf{r}_i(t-\delta t)+\frac{\delta t^2}{m_i} \textbf{f}_i(t)
\end{equation}
   and inserting the squared expression for  $\textbf{r}_i(t+\delta t) -\textbf{r}_i(t)$ in  Eq. (25), the change in kinetic energy
   is
\begin{equation}
	\delta K_{\textrm{D}}= \sum_i^N [ (\textbf{r}_i(t)-\textbf{r}_i(t - \delta t))\textbf{f}_i(t) +\frac{\delta t^2}{2m_i} \textbf{f}_i(t)^2].
\end{equation}

The energy invariance in Newton's discrete dynamics is expressed by Eqn. (24),  and  Eq. (28) as \cite{Toxvaerd2023}
\begin{equation}
	\delta E_{\textrm{D}}=	\delta U_{\textrm{D}}+\delta K_{\textrm{D}}=0.
\end{equation}
\section*{Acknowledgements}
The cooperation over many years with  my dear friend Luis F. Rull is gratefully acknowledged. 


\begin{thebibliography}{99}
	\bibitem{Newton1687} Newton, I.,1687, \textit{PHILOSOPHI\AE \ NATURALIS PRINCIPIA MATHEMATICA.} \textit{LONDINI}, \textit{Anno} MDCLXXXVII.
\bibitem{Toxvaerd2023} Toxvaerd, S., 2023,\textit{ Comprehensive Computational Chemistry} $ \textbf{3}$, 329 (Elsevier, Amsterdam, 2023).	
\bibitem{Toxvaerd1994}  Toxvaerd, S. 1994 Phys. Rev. E, {\bf 50}, 2271 . 
\bibitem{Toxvaerd2012}  Toxvaerd, S., Heilmann, O. J., and J. Dyre, J. C., 2012 \textit{J. Chem. Phys.}  {\bf 136}, 224106.
\bibitem{Verlet1967} Verlet, L., 1967, \textit{Phys. Rev.},  {\bf 159}, 98.
\bibitem{Toxvaerd2023appendix} See Reference No. 2, Appendix.
\bibitem{Tildesley} Allen, M. P., Tildesley, D. J., 1987,  {\it Computer Simulation of Liquids}
 (Oxford Science Publications, Oxford, 1987).	
\bibitem{Frenkel} D. Frenkel, D., Smit, B., 2023,  {\it Understanding Molecular Simulation} (Academic, New York, 2023).
\bibitem{Toxvaerd2023a} Toxvaerd, S., 2024,\textit{ Phys. Rev. E, in press}.	
\bibitem{Hoover1985} Hoover, W. G., 1985,\textit{ Phys. Rev. A}, $\textbf{31}$, 1695.
\bibitem{Harish2021} Harish M. S., and Patra P. K., 2021,\textit{ Mol. Phys.}, $\textbf{47}$, 701.
\bibitem{Hoover1991} Hoover, W. G., 1991, \textit{Computational Statistical Mechanics} (Elsevier, Amsterdam, 1991). 
\bibitem{Toxvaerd2013}  Toxvaerd, S., 2013,  \textit{J. Chem. Phys.}, {\bf 139}, 224106.
\bibitem{Toxvaerd1991}  Toxvaerd, S., 1991, \textit{Mol. Phys.}, {\bf 72}, 159.
\bibitem{Hoover1990} Holian, B. L., De Groot, A. J., Hoover, W. G, and Hoover, C. G., 1990, \textit{ Phys. Rev. A}, $\textbf{41}$, 3592.
\bibitem{Frenkelchap6}   Frenkel, D., Smit, B., 2002,  {\it Understanding Molecular Simulation}
 Chapter 6 and Appendix L,  (Academic, New York, 2002). . 
\bibitem{units} Units in MD of LJ systems: lengts in unit of $\sigma$, energy unit $\epsilon$, time unit $t^*=\sigma\sqrt{m/\epsilon}$.
	Temperature is $k_{\textrm{B}}T/\epsilon$. The LJ forces are cutted and shifted at $r_{cut}/\sigma=2.5$, for cut-and shifted
		forces in MD:   Toxvaerd, S., and Dyre, J. C., 2012, \textit{J. Chem. Phys.}  {\bf 134}, 081102.
\bibitem{Watanabe2012} Watanabe, H., Ito, N., and Hu, C-K., 2012, \textit{ J. Chem. Phys.},   {\bf 136}, 204102.
	\textit{Mol. Phys.},  {\bf 87}, 1117.
\bibitem{Martyna1992} Martyna, G. L., Klein, M., and Tuckerman, M., 1992, \textit{ J. Chem. Phys.},  {\bf 97}, 2635.
\bibitem{Patra2014} Patra, P. K., Bhattacharya, B., 2014, \textit{ Phys. Rev. E},  {\bf 90}, 043304.		
\bibitem{Rahman1964} Rahman, A., 1964,\textit{ Phys. Rev.}, $\textbf{136}$, A 405.
\bibitem{Orel1993} Jane\v{z}i\v{c}, D., and Orel, B., 1993, \textit{J. Chem. Inf. Comput. Sci.}, $\textbf{33}$, 252.		
\bibitem{Goldstein}   Goldstein, H., 1980, {\it Classical Mechanics},( Addison-Wesley Press Second Ed. 1980), Chap. 1. 
\end{thebibliography}
\end{document}